\documentclass[aps,showpacs,amsmath,amssymb,nofootinbib,preprintnumbers]{revtex4}
\usepackage{graphicx}
\usepackage{color}

\def \be  {\begin{equation}}
\def \ee  {\end{equation}}
\def \bea {\begin{eqnarray}}
\def \eea {\end{eqnarray}}

\begin{document}

\preprint{ECTP-2010-12}

\title{Matter-Antimatter Asymmetry in the Large Hadron Collider}

\author{A.~Tawfik}


\affiliation{The Egyptian Center for Theoretical Physics (ECTP) of the MTI University, Al-Hadaba Al-Wusta, Al-Mukattam, Cairo, Egypt}

\date{\today}

\begin{abstract}
The matter-antimatter asymmetry is one of the greatest challenges in the modern physics. The universe including this paper and even the reader him(her)self seems to be built up of ordinary matter only. Theoretically, the well-known Sakharov's conditions, which have been suggested more than 40 years ago, remain the solid framework explaining the circumstances that matter became dominant against the antimatter while the universe cools down and/or expands. On the other hand, the standard model for elementary particles apparently prevents at least two conditions out of them. In this work, we introduce a systematic study of the antiparticle-to-particle ratios measured in various $NN$ and $AA$ collisions over the last three decades. It is obvious that the available experimental facilities turn to be able to perform nuclear collisions, in which the matter-antimatter asymmetry raises from $\sim 0\%$ at AGS to $\sim100\%$ at LHC, where the center-of-energy raises from few GeV to $7000$ GeV, respectively.  
Assuming that the final state of hadronization in the nuclear collisions takes place along the freezeout line, which is defined by a constant entropy density, various antiparticle-to-particle ratios are studied in framework of the hadron resonance gas (HRG) model.  
Implementing modified phase space and distribution function in the grand-canonical ensemble and taking into account the experimental acceptance, the ratios of antiparticle-to-particle over the whole range of center-of-mass-energies are very well reproduced by the HRG model. Furthermore, the antiproton-to-proton ratios measured by ALICE in $pp$ collisions is also very well described by the HRG model. It is likely to conclude that the LHC heavy-ion program will produce the same particle ratios as the $pp$ program implying the dynamics and evolution of the system would not depend on the initial conditions. The ratios of bosons and baryons get very close to unity indicating that the matter-antimatter asymmetry nearly vanishes at LHC. The chemical potential calculated at this energy strengthens the assumption of almost fully matter-antimatter symmetry at LHC energy. Some cosmological consequences are introduced. 
\end{abstract}



\maketitle

\section{Introduction}
The elimination of antimatter while leaving behind some matter known as baryogenesis is one of the most fundamental problems in the modern physics. The Big Bang and cosmological standard model assuming homogeneous and isotropic matter (antimatter) distribution are conjectured to create matter and antimatter equally. But we experience only protons, neutrons, and electrons in our universe, i.e, the baryon asymmetry of the universe (BAU) seems to be $\sim100\%$. The universe including ourselves is therefore an obvious evidence that matter and antimatter are not fully symmetric. There are various astrophysical and fundamental observations supporting such a conclusion. The materialistic objects under non-extreme conditions have baryons and/or leptons only. The ratio of $n_{net-baryon}$ to $n_{photon}$ has been measured in cosmic microwave background (CMB) era by WMAP are found very much small $\sim 10^{-10}$ \cite{wmap,Tawfik:2008cd}. In the whole solar system there is no evidence for the existence of antimatter. Even in ultra highest energy cosmic rays (UHECR), the number of antiprotons, for instance, is in a good (fully) agreement with production as secondary particles from nuclear collisions. The observed gamma burst flux does not indicate the existence of antimatter in admixture of galaxies. It is apparent that the recent results of Large Hadron Collider (LHC) on antiparticle-to-particle ratios show that the produced matter and antimatter are almost equal \cite{alice2010,Tawfik:2010pt}.   

Under extreme conditions, like hot and dense QCD of nuclear collisions, the antimatter can be created. In present work, a phenomenological survey of the matter-antimatter asymmetry measured in nuclear interactions is introduced. Baryon and boson yield ratios are studied. The collective properties of hot and dense QCD is one of main objectives of the heavy-ion program. The possible modification of the transport coefficients, like phase structure and effective degrees of freedom, with increasing energy and size of the interacting system likely provides fruitful tools to study the collective properties. Furthermore, such  modifications seem to comprehensively characterize the particle and/or antiparticle production itself \cite{jeon}.
A universal description for particle and/or antiparticle production at different center-of-mass-energies of nucleon-nucleon ($NN$) collisions $\sqrt{s_{NN}}$ (hereafter simply $\sqrt{s}$) can be achieved by studying the dynamical fluctuations \cite{rajan,jeon} and multiplicities of particle yield ratios~\cite{Sollfrank:1999cy,Cleymans:1998yf}. The particle yield ratios are not only able to determine the freezeout parameters, temperature $T$ and chemical potential $\mu$, but also eliminate - to a very large extend - the volume fluctuations and the dependence of the freezeout surface on the initial conditions. 
On the other hand, the fluctuations and multiplicities of particle ratios in the  particle (antiparticle) production in nuclear collisions are also very much essential regarding to examining the statistical/thermal models \cite{giorg,Karsch:2003vd}, characterizing the chemical/thermal equilibration \cite{pequil} and finally search for unambiguous signals for the creation of the quark-gluon plasma (QGP) \cite{qgpA,qgpB}. 

The freezeout parameters ($T$ and $\mu$) can be determined in thermal models by combining various fluctuations and multiplicities of integrated particle (antiparticle) ratios~\cite{Sollfrank:1999cy,Cleymans:1998yf}. The search for common properties of these parameters at different center-of-mass-energies has a long tradition~\cite{Cleymans:1999st,Braun-Munzinger:1996mq}. Different models have been suggested so far~\cite{Cleymans:1999st,Braun-Munzinger:1996mq,Magas:2003wi,Tawfik:2005qn,Tawfik:2004ss,Tawfik:2005qh}. In present work, constant entropy density $s$ normalized to $T^3$ \cite{Tawfik:2004ss,Tawfik:2005qh} is used. This universal condition obviously implies that the hadronic matter is characterized by constant degrees of freedom. To illustrate this, the quantity $s/T^3$ can be scaled by $\pi^2/4$, which reflects the phase space. It results in the overall {\it effective} degrees of freedom of free gas~\cite{Cleymans:2005km}. It is worthwhile to mention that the phase transition (confined hadrons $\rightarrow$ deconfined QGP) itself is accompanied by a drastic decrease of the degrees of freedom \cite{Karsch:2003vd}.

Using grand--canonical ensembles and modified phase space in the hadron resonance gas (HRG) model makes it possible to fairly describe different features of the experimental data \cite{Tawfik:2010uh,Tawfik:2010aq} including particle multiplicities, dynamical fluctuations of various particle ratios \cite{tawPS} and thermodynamics of the strongly interacting system below the critical temperature $T_c$. With reference to the scope of present work, it has been noticed that the fluctuations and multiplicities over the whole range of $\sqrt{s}$ exhibit a non-monotonic behavior \cite{Tawfik:2010uh,Tawfik:2010aq}.

In heavy-ion collisions, the enhanced antiparticles are conjectured as indicators for the formation of deconfined QGP \cite{antib1}, whereas the possible annihilation might suppress such an enhancement \cite{antib2}. This would explain why the values of antiparticle-to-particle ratios in $pp$ collisions are higher than in heavy-ion collisions. Therefore, the initial conditions and formation time can be reflected by the surviving antiparticle. Decelerating or even stopping of incident particle (projectile) and its break up in inelastic collisions have been discussed in literature \cite{pbreak}. Therefore, the  antiparticle-to-particle ratios can be used to study particle (or antiparticle)  transport and production and therefore would have significant cosmological and astrophysical consequences. By mentioning astrophysics, it seems in order now to recall that the $n_{\bar{p}}/n_p$ ratios among others have been calculated and also observed in the cosmic rays revealing essential details on astrophysical phenomena \cite{antip_cr}. In present work, various antiparticle-to-particle ratios are calculated in the HRG model and compared with the heavy-ion collisions at $\sqrt{s}$ ranging from AGS to LHC, few GeV to couple TeV, respectively. Thus, the ratios and their description in the thermal model seem to provide fruitful tools to study the evolution of the matter-antimatter asymmetry with changing energy. In section \ref{sec:2}, the methods are introduced. Section \ref{sec:rslts} is devoted to the results and their discussion. Some cosmological and astrophysical consequences are discussed in section \ref{sec:astrophys}. The conclusions are listed in section \ref{sec:conls}.

\section{Methods}
\label{sec:2}

The antiparticle-to-particle ratios over the whole range of the center-of-mass-energy $\sqrt{s}$ are studied in framework of a thermal model.  
The hadron resonances treated as a free gas~\cite{Karsch:2003vd,Karsch:2003zq,Redlich:2004gp,Tawfik:2004sw,Taw3} are
conjectured to add to the thermodynamic quantities in the hadronic phase of heavy-ion collisions. This
statement is valid for {\it free} as well as {\it strong} interactions between the
hadron resonances themselves. It has been shown that the thermodynamics of strongly interacting
system can be approximated to an ideal gas composed of hadron
resonances~\cite{Tawfik:2004sw,Taw3,Vunog}. Also, it has been shown that the thermodynamics of strongly interacting system can be approximated to an ideal gas composed of hadron resonances with masses $\le 1.8~$GeV~\cite{Tawfik:2004sw,Taw3,Vunog}. The heavier constituents, the smaller thermodynamic quantities. The main motivation of using the Hamiltonian is that it contains all relevant degrees of freedom of {\it confined} and {\it strongly} interacting matter. It implicitly includes the interactions that result in the formation of {\it new} resonances. In addition, HRG model is used to provide a quite satisfactory description of particle production and collective properties in heavy-ion collisions \cite{Tawfik:2005qn,Tawfik:2010uh,Tawfik:2010aq,tawPS,Redlich:2004gp,Karsch:2003zq,Tawfik:2004sw,Taw3,Vunog}. 

At finite temperature $T$, strange $\mu_S$ and isospin  $\mu_{I_3}$
and baryochemical potential $\mu_B$, the partition function of one single particle and one single antiparticle reads \cite{Tawfik:2010uh,Tawfik:2010aq}
\bea
\ln Z_{gc}(T,{\cal V},\mu)&=& \sum_i\pm \frac{g_i}{2\pi^2}\,{\cal V}\int_0^{\infty} k^2 dk\; \left\{\ln\,\left(1\pm \gamma\,{\cal Q}\, \exp\left[\frac{\mu_i-\epsilon_i(k)}{T}\right]\right) + \ln\,\left(1\pm \gamma\,{\cal Q}\, \exp\left[\frac{-\mu_i-\epsilon_i(k)}{T}\right]\right)\right\}, \label{eq:zTr2}
\eea
where $\varepsilon_i(k)=(k^2+m_i^2)^{1/2}$ is the $i$-th particle's dispersion relation (single-particle energy) in equilibrium and $\pm$ stands for bosons and fermions, respectively. $g$ is the spin-isospin degeneracy factor and
$\gamma\equiv\gamma_q^n\gamma_s^m$ stand for the quark phase
space occupancy parameters, where $n$ and $m$ being the number of light and
strange quarks in the hadron of interest, respectively. $\mu$ is the chemical potential and $\lambda=\exp(c,\mu/T)$ is the fugacity factor. Multiplying $\mu$ by the corresponding charge $c$, like baryon, light and strange quark numbers, etc., makes it possible to count for particle and antiparticle. Since no phase transition is conjectured in HRG model (its validity is limited below $T_c$), summing over all hadron resonances results in the final thermodynamics in the hadronic phase. 

The generic parameter ${\cal Q}(\vec{x},\vec{k})$ is conjectured to reflect the change in phase-space, when the hadronic degrees of freedom are replaced by partonic ones and vice versa. The parameter $\gamma=\exp(-\alpha)$ can be interpreted as a measure for the non-extensivity, where $\alpha$ will be elaborated below. 
It gives the averaged occupancy of the phase space relative to
equilibrium limit. On the other hand, when assuming finite time evolution of the system, then $\gamma_i$ can be seen as a ratio of the change in the particle numbers before and after the chemical freeze-out, i.e. $\gamma_i=n_i(t)/n_i(\infty)$. The chemical freezeout turns to be defined as the time scale, at which no longer particle production is possible and the collisions become entirely elastic. 
To complete this picture, it is worthwhile to mention that in case of phase transition, $\gamma_i$ is expected to be larger than unity. The reasons are the large degrees of freedom, weak coupling and expanding phase space above the critical temperature to QGP.

The parameter $\alpha$ is a Lagrange multiplier in the entropy maximization. Its  physical meaning is a controller over the number of particles in the phase space, i.e, acting as chemical potential, $\alpha=\ln{\cal E}-\ln T-\ln N$.
It also combines intensive variables, $T$ and $N$ with an extensive one ${\cal E}=\sum_i^n g_i\epsilon_i$, where $\epsilon_i$ is energy density of $i$-th cell in the phase space of interest. 
The most probable state density can be found by the Lagrange multipliers, where one of them, $\alpha$, has been expressed in term of the second one, $\beta\equiv 1/T$, and the occupation numbers of the system. Apparently, $\alpha$ gives how the energy ${\cal E}$ is distributed in the microstates of the equilibrium system and therefore, can be understood as another factor controlling the number of occupied states at the microcanonical level.

The quark chemistry is given by relating the {\it hadronic} chemical potentials and
$\gamma$ to the quark constituents. $\gamma\equiv\gamma_q^n\gamma_s^m$ with
$n$ and $m$ being the number of light and strange quarks,
respectively. $\mu_B=3\mu_q$ and $\mu_S=\mu_q-\mu_s$, with $q$ and $s$ are
the light and strange quark quantum number, respectively. The
baryochemical potential for the light quarks is
$\mu_q=(\mu_u+\mu_d)/2$. The iso-spin
chemical potential $\mu_{I_3}=(\mu_u-\mu_d)/2$. Therefore, vanishing $\mu_{I_3}$ obviously means that light quarks are conjectured to be degenerate and $\mu_q=\mu_u=\mu_d$. The strangeness chemical potential, $\mu_S$, is calculated in HRG model as a function of $T$ and $\mu_B$. At finite $T$ and $\mu_B$,  $\mu_S$ is calculated under the condition of the overall strangeness conservation. 

Absolving chemical freezeout means that the hadrons finally decay to stable hadrons or resonances
\bea \label{eq:n2}
n_i^{final} &=& n_i^{direct} + \sum_{j\neq i} b_{j\rightarrow i} \; n_j,
\eea
where $b_{j\rightarrow i}$ being decay branching ratio of $j$-th hadron resonance
into $i$-th {\it stable} particle or antiparticle. The channel in which the particle and its antiparticle simultaneously appear is not taken into account. Such a channel of mixture decay seems to play an essential role in describing dynamical fluctuations of particle (antiparticle) yield ratios \cite{Tawfik:2004ss,Tawfik:2005qh,Tawfik:2005qn,Tawfik:2010uh,Tawfik:2010aq,tawPS}.

  
In grand-canonical ensemble, the particle number density is no longer constant. In an ensemble of $N_r$ hadron resonances,
\bea 
\label{eq:n1} 
\langle n\rangle &=& \sum_i^{N_r} \frac{g_i}{2\pi^2} \int_0^{\infty} k^2 dk \left\{
\frac{\lambda_i\,e^{-\epsilon_i(k)/T}}{({\cal Q}\,\gamma)^{-1}\pm \lambda_ie^{-\epsilon_i(k)/T}} \; +
\frac{\lambda_i^{-1}\,e^{-\epsilon_i(k)/T}}{({\cal Q}\,\gamma)^{-1}\pm \lambda_i^{-1}e^{-\epsilon_i(k)/T}} \right\}.
\eea

In carrying out the calculations given in present work, full grand-canonical statistical
sets of the thermodynamic quantities have been used. Corrections due to van~der~Waals
repulsive interactions have not been taken into account~\cite{Tawfik:2004sw}.
Seeking for simplicity, one can - for a moment - assume the Boltzmann approximations, especially at ultra-relativistic energies, in order find out the parameters, on which $\bar{n}/n$ ratios depend. Then, at finite isospin fugacity $\lambda_{I_3}$, the ratios of antiproton-to-proton and antikaon-to-kaon, respectively, can be approximated as
\bea
\frac{n_{\bar{p}}}{n_{p}}  &\simeq&  
      \frac{\lambda_{\bar{p}}}{\lambda_{p}} 
      = \left(\lambda_{\bar{u}}^2\; \lambda_{\bar{d}}\right)^2, \label{naptip-p}\\
\frac{n_{K^{-}}}{n_{K^+}}  &\simeq&  \frac{\lambda_{K^{-}}}{\lambda_{K^+}} =
      \left(\lambda_{\bar{u}}\; \lambda_{S}\right)^2. \label{nkm-kp}
\eea
These two expressions can be used to calculate the light and strange quarks chemical potentials at various energies using the {\it experimentally} measured ratios. As discussed above, both baron and strange potentials are correlated to assure an overall strangeness conservation.  

\section{Results and Discussion}
\label{sec:rslts}

As mentioned above, the formation of deconfined QGP strongly depends on the initial conditions in the early stages in nuclear interactions, where the particle (antiparticle) number transport (deceleration of the incoming projectile) or even its entire stopping is assumed to be achieved \cite{nb1}. Therefore, it is widely conjectured that such deconfined matter would not be produced, especially in $pp$ collisions at low energies. On the other hand, the degrees of transport likely affect the overall dynamical evolution of the system. The initial parton equilibrium \cite{ipd}, thermal/chemical equilibrium \cite{tce}, time evolution  \cite{tev} and the final particle production \cite{pp} are affected as well. 

\begin{figure}[thb]
\includegraphics[width=8.cm,angle=-90]{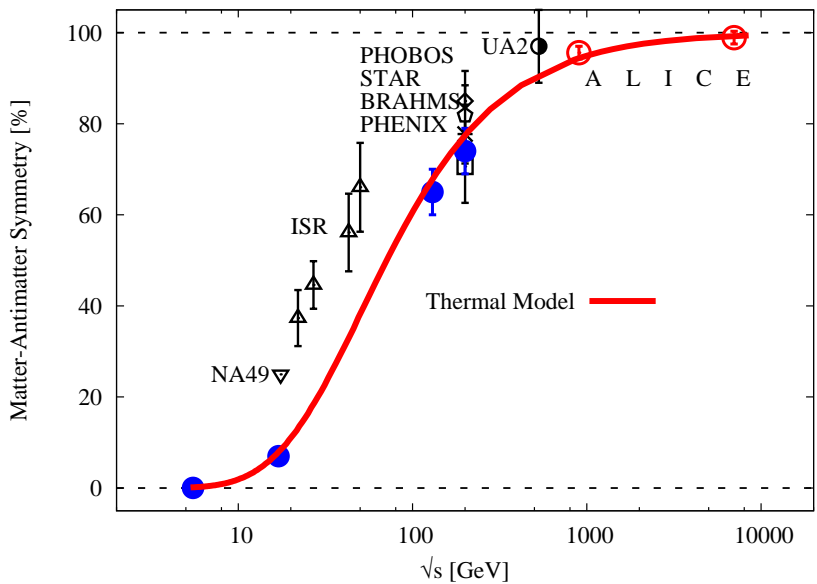}
\caption{\normalsize $n_{\bar{p}}/n_p$ ratios depicted in the whole available range of $\sqrt{s}$. Open symbols stand for the results from various $pp$ experiments (labeled). The heavy-ion results from AGS, SPS and RHIC, respectively, are drawn in solid symbols. The solid curve represents results from thermal HRG model at ${\cal Q}=0.5$ and $\gamma=1.0$, which apparently perfectly describes all heavy-ion collisions, besides the ALICE $pp$ results. At LHC energy, the ratios are very much close to unity indication almost $100\%$ matter-antimatter symmetry.} 
\label{fig:1a}
\end{figure}

In Fig. \ref{fig:1a}, the baryonic $n_{\bar{p}}/n_p$ ratios are given in dependence on center-of-mass-energy $\sqrt{s}$ ranging from AGS to LHC, few GeV to couple TeV, respectively. The $pp$ collisions (empty symbols) are compared with the heavy-ion collisions (solid symbols). In the nucleus-nucleus ($pp$) collisions, the spatial and time evolution of the system is too short to assure initial conditions required to drive the parton matter into deconfined QGP. We notice that $n_{\bar{p}}/n_p$ ratios up to the RHIC energy are smaller than $100\%$ indicating an overall excess of proton $n_p$ over antiproton $n_{\bar{p}}$. 

As mentioned above, the HRG model has been successfully used to describe the heavy-ion collisions and the thermodynamics of lattice QCD as well \cite{Karsch:2003vd,Tawfik:2005qn,Tawfik:2004ss,Tawfik:2005qh,Karsch:2003zq,Redlich:2004gp,Tawfik:2004sw}. The collective flow of the {\it strongly} interacting matter in heavy-ion collisions makes it obvious that the HRG model underestimates the $n_{\bar{p}}/n_p$ ratios measured in $pp$ collisions, especially at low energies. On the other hand, the ratios from the heavy-ion collisions \cite{star1} are very well reproduced by the HRG model. In both types of collisions, there is a dramatic increase in the $n_{\bar{p}}/n_p$ ratio \cite{SPS1,AGS1,alice2010}. 
At LHC energy \cite{alice2010}, while the differences between $n_{\bar{p}}/n_p$ ratios of $pp$ and heavy-ion collisions apparently {\it almost} disappear, their values approach $100\%$, i.e. {\it nearly} $0\%$ matter-antimatter asymmetry. 

The solid line gives the results from the HRG model, section \ref{sec:2}. No fitting has been utilized. Over the whole range of $\sqrt{s}$, the two parameters $\gamma$ and ${\cal Q}$ remain constant, $1.0$ and $0.5$, respectively. HRG obviously describes very well the heavy-ion results. Also, the ALICE $pp$ results are reproduced by means of HRG model. Therefore, the solid line seems to be a universal curve describing the behavior of antiproton-to-proton ratios, while $\sqrt{s}$ raises from few GeV to couple TeV. Furthermore, the LHC results indicate an {\it almost} vanishing net-baryonic matter. 

\begin{figure}[thb]
\includegraphics[width=8.cm,angle=-90]{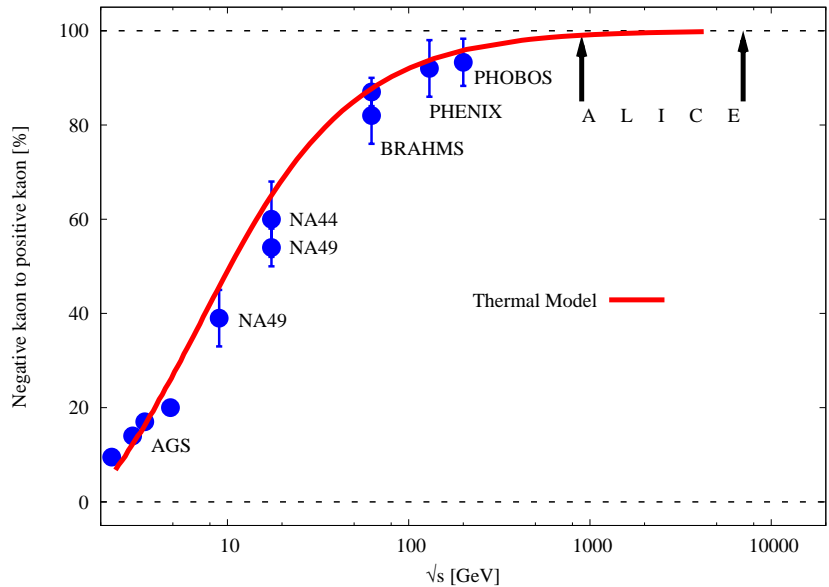}
\caption{\normalsize $n_{K^-}/n_{K^+}$ ratios given in the whole available range of $\sqrt{s}$ in heavy-ion collisions of AGS, SPS and RHIC, respectively. The solid curve represents HRG results. At RHIC energy, the ratios reach $93\%$, whereas at LHC energy very much close to unity indication almost $100\%$ antiboson-boson symmetry. The values predicted at ALICE energies are marked by upwards arrows.} 
\label{fig:1b}
\end{figure}

In Fig. \ref{fig:1b}, the ratios $n_{K^-}/n_{K^+}$ are given in dependence on $\sqrt{s}$ ranging from AGS to LHC, few GeV to couple TeV, respectively. Once again in this bosonic antiparticle-to-particle ratios, there is a drastic raise enhancing the accessibility of negative against positive kaons with increasing $\sqrt{s}$ \cite{brahms01,phobos01,phenix01,kaos01,sis01}. At RHIC energies, the symmetry of these bosons approaches $\sim93\%$, which is obviously higher than the baryonic antiproton-proton in Fig. \ref{fig:1a} ($\sim73\%)$. Same behavior is also observed at lower energies. At the same energy, the values of $n_{K^-}/n_{K^+}$ are higher than the values of $n_{\bar{p}}/n_p$ implying that bosons are relatively much easier to be accessed than the baryons. Comparing the two figures illustrates these differences. 
In $pp$ collisions, there are few measurements at low energy \cite{na49-01}. Therefore, the comparison with the heavy-ion collisions would be incomplete. At LHC energy, the ratio gets very much close to $100\%$ indicating a nearly {\it fully} antiboson-to-boson symmetry.

\section{Cosmological and astrophysical consequences}
\label{sec:astrophys}
  
\begin{figure}[thb]
\includegraphics[width=8.cm,angle=-90]{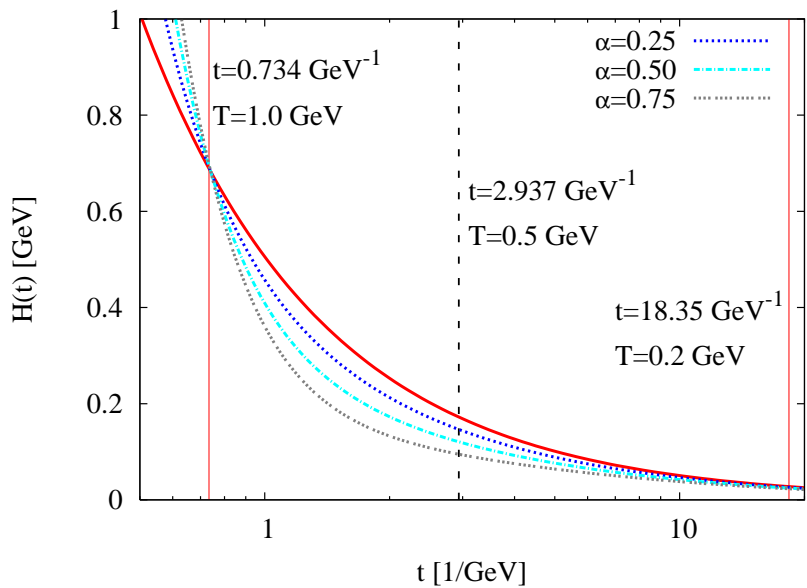}
\caption{\normalsize The time evolution of the Hubble parameter $H(t)$ at
 different values of the perturbative parameter for bulk viscosity, $\alpha$. The vertical lines determine the time $t$ and temperature $T$ boundaries of the QCD era in early universe which can be assigned to LHC energy.} 
\label{fig:3a}
\end{figure}

In an isotropic and homogeneous background, the volume of the universe seems to be directly related to the scale parameter $a(t)$, where $t$ is the co-moving time. Implementing barotropic equation of state for the background matter makes it possible to calculate - among others - the Hubble parameter $H(t)=\dot a(t)/a(t)$. Dot refers to the first derivative with respect to time $t$. Focusing the discussion on QCD era of early universe, which likely turns to be fairly accessible in the high-energy experiments, the equation of state is very well defined, experimentally. In Ref. \cite{Tawfik:2010ht}, a viscous equation of state for QGP matter has been introduced. Different solutions for the evolution equation of $H$ have been worked out. For simplicity we can assume that the deviation of $H$ in viscous background matter from the one in the ideal matter can perturbatively be given as \cite{Tawfik:2010pm}
\begin{equation}\label{eq:6}
H(t)=H_0(t)+\alpha f(t),
\end{equation}
where $H_0(t)$ is the Hubble parameter corresponding to a vanishing bulk viscosity. The parameter $\alpha$ will be given in Eq. (\ref{eq:xixi}). In this treatment, it is taken a perturbative variable. The barotropic equation of state of viscous QCD matter (QGP matter) are 
\bea\label{13an}
P &=& \omega \rho, \label{eq:pp}\\
T &=& \beta \rho^r,\label{eq:tt} \\
\xi &=& \alpha \rho, \label{eq:xixi}
\eea
besides an expression for the relaxation time. 
The quantities $P$, $\rho$, $\xi$ and $T$ are pressure, energy density, bulk viscosity and temperature, respectively. The parameters $\omega$, $\beta$, $\alpha$ and $r$ are to be determined from the high-energy experiments and/or the lattice QCD simulations \cite{Karsch:2003vd,Karsch:2003zq,Redlich:2004gp,Tawfik:2004sw,Taw3}. 
The function $f(t)$ has been evaluated \cite{Tawfik:2010pm}. The resulting $H(t)$ is depicted in Fig. \ref{fig:3a}. The vertical lines define the temperatures available to the QGP matter according to heavy-ion experiments and lattice QCD simulations and simultaneously the time (in 1/GeV units) according to the Big Bang theory. By doing this, the volume $V$ and scale factor $a$ in the early universe can be roughly estimated.

The ratio of baryon density asymmetry  to photon density, $\eta$, has been measured in WMAP data \cite{wmap}. Then $n_{\bar{p}}-n_p$ from ALICE experiment at $7\,$GeV can be user to give an estimation of the photon number density.
\bea
4.4\times 10^{4}\; < & n_{\gamma} & <\; 7.7\times 10^{4}. \label{eq:gammaALICE1}
\eea
When introducing a model for the volume of the early universe as given in Fig. \ref{fig:3a} at energies equal to top ALICE, $7\,$TeV, the photon number, $N_{\gamma}=n_{\gamma}\,V$, can be calculated. It is obvious that the resulting $N_{\gamma}$ will be extraordinarily large indicating dominant radiation against all other equations of state in this era where matter and antimatter are supposed to be almost symmetric. In CMB era, where $T\approx 2.73\,$K, the photon number density $n_{\gamma} \approx 411.4 (T/2.73 \mathtt{K})\,$ cm$^{-3}$.

\begin{figure}[thb]
\includegraphics[width=8.cm,angle=-90]{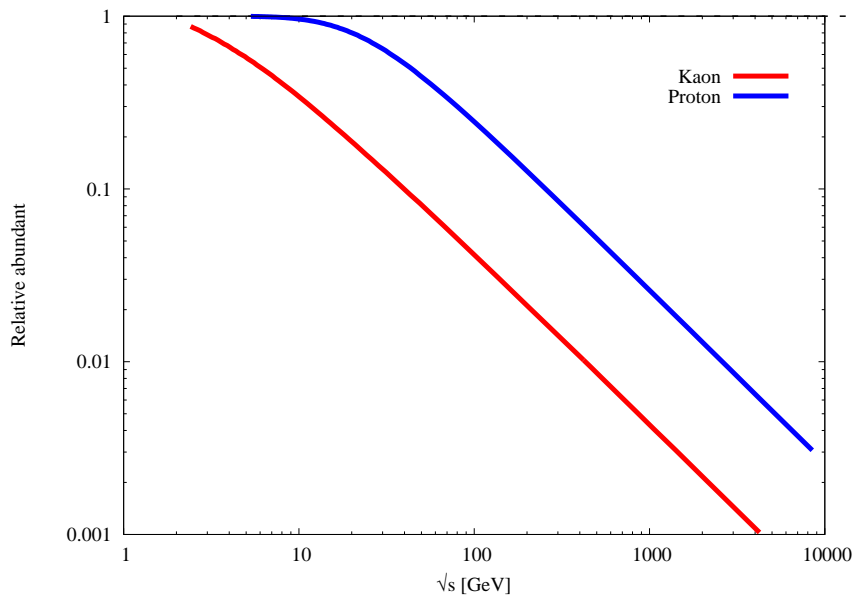}
\caption{\normalsize The relative abundant of kaon and proton yields as functions of $\sqrt{s}$ in log-log chart. Net boson and proton relative to their summation  almost linearly decrease with increasing $\sqrt{s}$. Apparently, kaon relative abundant is by about one order of magnitude faster than the proton. } 
\label{fig:4a}
\end{figure}

In a co-moving volume $V\sim a^3(t)$, the number density of non-interacting photons likely remain constant. Therefore
\bea
n_{\gamma} &\sim& 1/a^3(t).
\eea
Nevertheless, the decrease of temperature, when the universe expands, would affect the quantity $a^3(t)\,n_{\gamma}$. As a consequence, $a^3(t)\,n_{\gamma}$ seems to change in the expanding universe. The values in CMB and in Eq. (\ref{eq:gammaALICE1}), where the latter should be related to an era characterized by ALICE temperature, can be taken as an indicator for varying gamma number with cooling down the early universe. There is a conserved quantity, namely the entropy density $s$. In a {\it perfectly} closed system like the universe, $s$ likely remain unchanged.  It can be deduced from the first-law of thermodynamics and Eq. (\ref{eq:zTr2}) \cite{Tawfik:2010ht}. At vanishing chemical potential, Eq. (\ref{eq:pp}) and (\ref{eq:tt}), which are valid for the QGP matter, lead to, 
\bea
s(T) &=& \frac{P(T)+\rho(T)}{T} = \frac{\omega+1}{\beta}\;\rho(T)^{1-r},
\eea
implying that $s(T)$ is related to $\rho(T)^{1-r}$, where $\omega=0.319$, $\beta=0.718\pm 0.054$ and $r=0.23\pm 0.196$. At low energy, where the dominant degrees of freedom are given by all hadron resonances, Eq. (\ref{eq:zTr2}). The entropy density can be derived from Eq. (\ref{eq:zTr2}). Doing this, we can study baryon and boson relative abundant $(n(T)-\bar{n}(T))/(n(T)+\bar{n}(T))$. In Fig. (\ref{fig:4a}), HRG results for the relative abundant as a function on $\sqrt{s}$ are given for both spices. We notice that the kaon relative abundant is by about one order of magnitude aster than the proton. It is obvious that the abundances of light elements ($^7$Li, $^4$He, $^3$He and $^2$H) produced in the early universe are sensitive indicators of the baryonic number density \cite{lightel}. 

\section{Conclusions}
\label{sec:conls}

So far, it can be concluded that the future LHC heavy-ion program likely will produce antiparticle-to-particle ratios similar, when not entirely identical, to the ones measured in this current $pp$ program. Secondly, the ratios obviously run very close to unity implying almost vanishing matter-antimatter asymmetry. Some astrophysical consequences of these results are given in section \ref{sec:astrophys}. On the other hand, it can also be concluded that the thermal models including the HRG model seem to be able to perfectly describe the hadronization at very large energies and the condition deriving the chemical freezeout at the final state of hadronization, the constant degrees of freedom or $S(\sqrt{s},T)=7\,(4/\pi^2)\,V\,T^3$, where $V$ (or ${\cal V}$ as in Eq. (\ref{eq:zTr2})) is the volume, seems to be valid at all center-of-mass-energies raging from AGS to LHC. Furthermore, at very high energy, the initial conditions likely do not affect the particle (antiparticle) production so that the ratios in $NN$ collisions equal the ratios in $AA$ collisions. The rate and amount of annihilation of antiparticle obviously do not differ in both types of collisions. At the freezeout boundaries, the particle (antiparticle) production seems to be well described by the degrees of freedom. At LHC energy, either antiboson-to-boson or antibaryon-to-baryon ratios are equal implying nearly vanishing net-bosons and net-baryons. The vanishing net-bosons are much faster than the vanishing net-baryons.    

\begin{figure}[thb]
\includegraphics[width=8.cm,angle=-90]{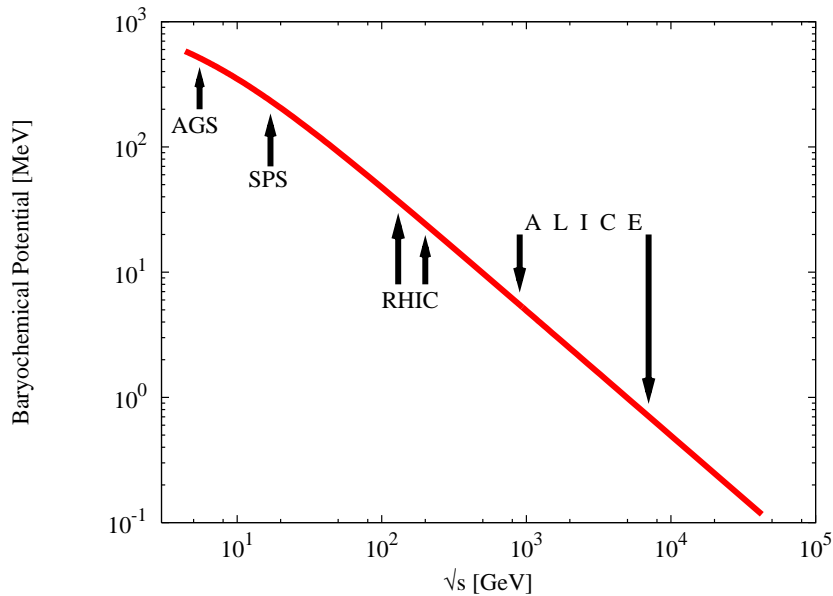}
\caption{\normalsize The baryochemical potential $\mu_B$ is depicted versus the center-of-mass energy $\sqrt{s}$ in a log-log chart. With increasing $\sqrt{s}$, $\mu_B$ decreases almost linearly indicating that the higher collision energies, the smaller energy required to insert or remove (create or annihilate) {\it new} particles.} 
\label{fig:2a}
\end{figure}

The expression (\ref{naptip-p}) and (\ref{nkm-kp}) can be used to calculate the antiquark and strange chemical potentials, respectively. Seeking for simplicity, we assume vanishing isospin chemical potential $\mu_{I_3}$, i. e, degenerate light quarks 
\bea
\frac{n_{\bar{p}}}{n_{p}}  &\simeq&  
      \exp(-6\mu_{\bar{q}}/T_{ch}), \label{naptip-p2} \\
\frac{n_{K^{-}}}{n_{K^{+}}} &\simeq& \exp(-2\mu_{\bar{u}}/T_{ch})\;\exp(2\mu_{s}/T_{ch}), \label{nkm-kp2}
\eea
where $T_{ch}$ is the freezeout temperature. Substituting the value of $n_{\bar{p}}/n_p$ measured by ALICE experiment at $7\,$GeV and $T_{ch}=0.175\pm0.02\,$GeV \cite{alice2010} in Eq. (\ref{naptip-p2}) results is
\bea
\mu_{\bar{q}} &\simeq& 0.179\;\;\mathtt{MeV}, \label{naptip-p3}.
\eea
Then, extracting $n_{K^-}/n_{K^+}$ at the same energy, $7\,$TeV, from Fig. \ref{fig:1b} and substituting it in Eq. (\ref{nkm-kp2}) leads to
\bea
\mu_s &\simeq& 0.0038 \;\;\mathtt{MeV}. \label{nkm-kp3}
\eea
The evolution of baryochemical potential $\mu_B$ for hadronic baryons with increasing $\sqrt{s}$ is illustrated in Fig. \ref{fig:2a}. The larger $\sqrt{s}$, the smaller $\mu_B$. The relations between $\mu_B$ and the quark chemistry is given in section \ref{sec:2}. At LHC energy, $\mu_B$ is very small indicating a very small difference between particle and antiparticle. This situation likely fairly simulates the early universe during the QCD era \cite{Tawfik:2010pm,Tawfik:2010mb,Tawfik:2010ht,Tawfik:2010bm,Tawfik:2009nh,Tawfik:2009mk}.

The matter-antimatter asymmetry is one of the greatest mysteries in modern physics. The well-known Sakharov's conditions \cite{sakhrv}; baryon number violation and C and CP violation besides out-of-equilibrium processes; are conjectured to give a solid framework to explain why should matter become dominant against antimatter with the universe cooling-down of expansion. Recently, a fourth condition has been suggested; $b$-$l$ violation, where $b$ and $l$ being baryon and lepton quantum number, respectively. The first condition means a different interactions of particles and antiparticles. The second condition means nonconservation of the baryonic charge. These two conditions are not given in the standard model, where both $b$ and $l$ are conserved, explicitly. Non-perturbatively, $b+l$ can be violated through sphalerons, for instance, which drive the two-direction conversion $b\leftrightarrow l$ with a certain number of selection rules \cite{sphl}. That the heavy-ion program seems to give systematic tools to survey the evolution of matter-antimatter asymmetry from $\sim 0\%$ to $\sim 100\%$ in the nuclear interactions would make it possible to devote such experimental facilities to study the evolution of Sakharov's conditions with the center-of-mass-energy $\sqrt{s}$. 

Phenomenologically, we need to find explanation for the different relative abundances of bosons and baryons and for the very small charge asymmetry during early universe. This would be the first step in phenomenological studies of Sakharov's conditions. Although all effective models including thermal ones and the lattice QCD thermodynamics apparently assume an equilibrium system, thermal out-of-equilibrium processes have to be implemented.

\section{Acknowledgment}

This work is partly supported by the Science and Technology Development Fund (STDF) under the German-Egyptian Scientific Projects (GESP) grant. The author likes to thank ALICE collaboration for providing the latest measurement of antiproton-to-proton ratios at $0.7$ and $9\,$TeV and Tim Schuster for enlightening discussions on measuring particle yields.

%
%

\end{document}